\documentclass [12pt]{article}
\usepackage{amssymb,amsmath}
\def\ddd{\displaystyle}
\def\C{{\mathbb C}}
\def\N{{\mathbb N}}

\def\R{{\mathbb R}}
\def\D{\overline{D}}
\def\A{{\cal A}}

\def\E{{\cal E}}
\def\P{{\cal P}}
\def\H{{\cal H}}

\def\epsilon{\varepsilon}
\def\phi{\varphi}
\def\kappa{\varkappa}
\def\theorem#1{\smallskip
{\scshape Theorem} {\bf #1}{\bf .}\hskip 8pt\sl}
\def\lemma#1{\smallskip
{\scshape Lemma} {\bf #1}{\bf .}\hskip 6pt\sl}
\def\epr{\smallskip\rm}
\def\wz{\thinspace}
\def\proof{P\wz r\wz o\wz o\wz f.\hskip 6pt}

\def\leq{\leqslant}
\def\geq{\geqslant}
\def\deg{\text{\rm deg}\,}
\def\im{\text{\rm Im}\,}
\def\re{\text{\rm Re}\,}

\def\K{\overline{K}}
\def\square{\vrule height6pt width6pt depth 0pt}

\begin {document}

\begin {center}
{\huge The resonance spectrum of the cusp map
in the space of analytic functions}

\vskip1truecm

\Large \rm
I.~Antoniou$^{1,2}$,
S.~A.~Shkarin$^{2,3,5}$ {\large and}
E.~Yarevsky$^{2,4}$
\end{center} \small

\vskip 0.3cm \noindent
$^1${Department of Mathematics, Aristotle University
of Thessaloniki, 54006, Greece}\hfill\break
$^2${International Solvay Institutes for
Physics and Chemistry, Campus Plaine ULB C.P.231, Bd.du
Triomphe, Brussels 1050, Belgium}\hfill\break
$^3${Moscow State University, Dept. of
Mathematics and Mechanics, Vorobjovy Gory, Moscow, 119899,
Russia} \hfill\break
$^4${Laboratory of Complex Systems Theory, Ins
titute
for Physics, St. Petersburg State University, Uljanovskaya
1, Petrodvoretz, St. Petersburg 198904, Russia} \hfill\break
$^5${Department of Mathematics, Wuppertal
University, 42119, Wuppertal, Gauss str. 20, Germany}

\smallskip

\tt\noindent {\bf e-mail:} \ eiarevsk@ulb.ac.be\quad
shkarin@math.uni-wuppertal.de\normalsize\rm

\begin {abstract}
We prove that the Frobenius--Perron operator $U$ of the cusp
map $F:[-1,1]\to[-1,1]$, $F(x)=1-2\sqrt{|x|}$ (which is
an approximation of the Poincar\'e section of the Lorenz
attractor) has no analytic eigenfunctions corresponding to
eigenvalues different from 0 and 1.
We also prove that for any $q\in(0,1)$ the spectrum of
$U$ in the Hardy space in the disk $\{z\in\C:|z-q|<1+q\}$
is the union of the segment $[0,1]$ and some finite or
countably infinite set of isolated eigenvalues of finite
multiplicity.
\end {abstract}

\newpage

\section {Introduction}

The so-called cusp map \cite{hemmer}
$$
F:[-1,1]\to[-1,1],\qquad F(x)=1-2\sqrt{|x|}
$$
is an approximation of the Poincar\'e section of the Lorenz
attractor~\cite{lorenz,attractor}. This map is ergodic~\cite{sinai}.
The unique absolutely continuous
invariant probability measure $\mu$ has density
$\rho(x)=(1-x)/2$ \cite{hemmer}.

The Frobenius--Perron operator (F.P.O.) $U$ of $F$ is
the adjoint of the Koopman operator $V$ \cite {lasota}
in the Hilbert space $L_2([-1,1],\mu)$:
$$
Vf(x)=f(F(x)),
$$
\begin{equation}
Uf(x)=\frac12\left(1-\frac{(1-x)^2}4\right)
f\left(\frac{(1-x)^2}4\right)+
\frac12\left(1+\frac{(1-x)^2}4\right)
f\left(-\frac{(1-x)^2}4\right).
\label {fro}
\end{equation}
The spectral analysis of the F.P.O. in different function spaces
is useful for the probabilistic approach to non-linear dynamics.
The spectrum of the F.P.O. known also as resonance spectrum gives
estimates on the decay of correlation functions,
see, e.g. \cite {lasota,baladi,ergodic}. The spectral decomposition
of the Koopman and Frobenius--Perron operators acquires meaning in
locally convex topological spaces and allows for probabilistic
prediction~\cite {ant42,ant76,ant94,ant143}.

The cusp map is not expanding \cite {baladi}.
In  \cite {gy,kauf,lust}, the following family of maps depending
analytically on the parameter $\epsilon\in(0,1/2]$ is introduced
and studied:
\begin {equation}
F_\epsilon : [-1,1] \rightarrow [-1,1], \qquad
F_\epsilon(x) = {1-\sqrt{1-4\epsilon(1-\epsilon-2|x|)} \over 2\epsilon},
{\rm \ \ for\ \ } \epsilon\in(0,1/2].
\label {familydefinition}
\end {equation}
This family consists of piecewise analytic expanding maps and has the
cusp map as the limit case for $\epsilon=1/2$.
For any map~(\ref{familydefinition}), the spectrum of the F.P.O.
in the space of $C^\infty$ functions consists
of a sequence of eigenvalues of finite multiplicity converging to 0,
and the corresponding eigenfunctions are analytic~\cite {ru3,baladi}.
The divergence of the eigenfunctions as the maps~(\ref{familydefinition})
approach the cusp map has been observed numerically~\cite {kauf,lust,we}.
These numerical results indicate that the F.P.O. (\ref{fro}) has
no analytic eigenfunctions corresponding to eigenvalues different from
$0$ or $1$. In the present
paper we give an analytic proof of this fact. Our result confirms the
reliability of the numerical works~\cite {kauf,lust,we}.

The spectral properties of the F.P.O. of piecewise analytic maps with
one neutral fixed point (i.e. fixed point with derivative
equal to 1; this is the point $x=-1$ for the cusp map) have been
addressed by Rugh \cite {rugh}. 
For each map satisfying certain properties, Rugh 
has constructed a map-dependent Banach space of functions, analytic
everywhere except at the neutral fixed point.
The spectrum of the Frobenius--Perron operator in this Banach space 
is the union of the segment $[0,1]$ and some isolated eigenvalues 
of finite multiplicity. However, Rugh's results do not specify the
spectrum of the F.P.O in the space of everywhere analytic
functions. Moreover, the cusp map does not satisfy the properties
of the class of maps considered by Rugh as a result of the cusp 
singularity. Nevertheless, we prove in this paper that for
any $q\in(0,1)$ the spectrum of the F.P.O.~(\ref{fro})
in the usual Hilbert Hardy space in the disk
$\{z\in\C:|z-q|<1+q\}$ is the union of the segment $[0,1]$
and a finite or countable set of isolated eigenvalues of
finite multiplicity. 

The paper is organized as follows. 
In Section~2, we summarize in Theorems~1--4
our results about spectra of the
operator $U$, and derive Theorems~1 and 2.
We prove auxiliary lemmas in Section~3, while 
the main proofs are presented in Sections~4 and~5.

\section {The spectrum of the F.P.O.}

We shall use the following notations.
$\A$ is the space of real-analytic functions
$f:[-1,1]\to\C$ and $\E$ is the space of entire functions
of one complex variable (endowed with their natural
topologies \cite{rob}).  $\P$ is the space of polynomials
of one complex variable. As usual in algebra, we assume
that the degree of zero polynomial is $-1$.

The spectral structure of the F.P.O. in the spaces $\A$ and
$\E$ is given by the following two theorems.

\theorem1 \\
The spectra of both operators
$U_\E=U\bigr|_\E:\E\to\E$ and $U_\A=U\bigr|_\A:\A\to\A$
coincide with the whole complex plane. \epr

\theorem2 \\
I. The point spectra of both operators $U_\E$ and $U_\A$
coincide with the two-point set $\{0,1\}$. The eigenvalue $1$ is
simple and the eigenvalue $0$ has infinite multiplicity
for both operators. \\
II. Let $f\in\A$. Then $Uf\equiv0$ if and only if
$f(x)\equiv(1+x)g(x)$ for all $x\in[-1,1]$, where
$g\in\A$ is odd. \\
III. Let $\lambda\in\C$, $f\in\A$, $n\in\N$ and
$(U-\lambda I)^nf=0$ $($where $I$ is the identity operator$)$.
\begin{itemize}
\item[{\rm (i)}] If $\lambda\notin\{0,1\}$ then $f=0$.
\item[{\rm (ii)}] If $\lambda=1$ then $f$ is constant.
\end{itemize}\epr

As we have already mentioned, Rugh's theorem~\cite {rugh} cannot be
applied for the cusp map. Moreover, one cannot use the technique
of Rugh's proof to prove a similar theorem for the cusp map.
Indeed, the key element of Rugh's proof is
a holomorphic function $\phi$, defined in an open simply
connected set, containing all points of the segment except
the neutral fixed point such that
$\phi\circ\xi\circ\phi^{-1}(z)\equiv z+1$, where $\xi$ is
a branch of the inverse map containing the neutral fixed
point ($\xi(z)=-(1-z)^2/4$ for the cusp map). Such a map
$\phi$ obviously does not exist if $\xi'$ vanishes somewhere
on the segment, which is the case for the cusp map since
$\xi'(1)=0$. Nevertheless, using similar ideas but quite
different technique, we prove
that $U$ has the spectral structure as in Rugh's theorem in
appropriate Hardy spaces. The definitions of the Hardy
spaces $H^2$ in the unit disk and in the upper half-plane
$\{z:\text{Im}\,z\geq0\}$ can be found e.g.  in
\cite {hoffman}, Chapters~3 and 8. These spaces are
separable Hilbert spaces. 

\theorem3 \\
Let $q\in(0,1)$ and $X$ be the Hardy space $H^2$ in
the disk $\{z\in\C:|z-q|<1+q\}$. Then $U$ is a bounded
operator on the Banach space $X$ and the spectrum of
$U\bigr|_{X}:X\to X$ is the union of the segment $[0,1]$
and a finite or countable set of isolated eigenvalues of
finite multiplicity. \epr

The proof of Theorem~3 is given in Section~5. We do not consider 
Hardy spaces $H^p$, $p\neq2$, although the Theorem~3
can be generalized for such spaces. The Hardy spaces
in any other disk or half-plane are the results of
appropriate linear change of variables applied to the
Hardy space in the unit disk or the upper half-plane.

For the sake of completeness we formulate here the following theorem
proved in \cite {we}.

\theorem4 \\
The spectrum of the operator
$U\bigr|_{X}:X\to X$, where $X$ is either $L_p([-1,1],\mu)$
$(1\leq p\leq+\infty)$ or $C^k[-1,1]$
$(k=0,1,\dots,\infty)$, is the closed unit disk.
The point spectrum of $U\bigr|_{X}$ is the set $\{z\in\C :|z|<1\}\cup \{1\}$.
The eigenspace corresponding to the eigenvalue $1$ is the one
dimensional space of constants.
The eigenspace corresponding to the eigenvalues $\{z: |z|<1\}$
is infinite dimensional. \epr

For the proof of Theorems~1 and~2 we need the following

\smallskip {\scshape Proposition.} \\ \sl
I. Let $\lambda\in\C\setminus\{0\}$, $g\in\E$, $f\in\A$ and
\begin {equation}
Uf(x)=\lambda f(x)+g(x),
\label {fe1}
\end {equation}
for all $x\in[-1,1]$. Then $f\in\E$.\\
II. If additionally $g\in\P$, then $f\in\P$ and
$\deg f\leq \deg g+1$.\\
III. If $g\in\P$ and $\deg g=4k+1$, $k=0,1,\dots$ then the
functional equation $(\ref{fe1})$ has no solutions in $\A$.
\epr

The proof of the Proposition is given in Section~4.

{\bf Proof of Theorem 1.} \
The number 0 is an eigenvalue of $U_\A$ and $U_\E$. For example
$Uf=0$ if $f(x)=(1+x)x$. Therefore, the number $0$ belongs to the
spectra of these operators.  Let
$\lambda\in\C\setminus\{0\}$. According to the Proposition,
the functional equation~(\ref{fe1}) has in the space $\P$
no solutions for $g$ with degree of the form $4k+1$.  Thus,
the functional equation $Uf(x)=\lambda f(x)+x$ has no
analytic solutions.  Therefore the function $g(x)=x$ does
not belong to the image of the operator $U_\A-\lambda I$
(and to the smaller image of $U_\E-\lambda I$) for any
$\lambda\in\C\setminus\{0\}$. Hence, operators
$U_\A-\lambda I$ and $U_\E-\lambda I$ are
non-invertible. Therefore the spectra of $U_\A$ and $U_\E$
coincide with the whole complex plane.

{\bf Proof of Theorem 2.} \
Let $z\in\C\setminus\{0\}$, $c\in\C$ and $f\in\A$.
The Proposition implies that
\begin{equation}
\text{if }\ Uf-zf\equiv c\text{\ \ then\ \ }f \text{\ \
is a constant}.
\label {con}
\end{equation}

Let $\lambda\in\C$, $\lambda\neq 0$, $f\in\A$ and $(U-\lambda I)^n f=0$.
Relation (\ref{con}) implies consequently that the functions
$h_1=(U-\lambda I)^{n-1}f$, $h_2=(U-\lambda I)^{n-2}f$,
$\dots$, $h_{k-1}=(U-\lambda I)f$, $h_k=f$ are constants. In
particular, $f\equiv a$ for some $a\in\C$.  Since $U1=1$,
we have that $0=(U-\lambda I)^nf=(1-\lambda)^n\cdot a$.  Therefore either
$\lambda=1$ or $f=0$.  This proves part III of the theorem.

Let now $f\in\A$. According to (\ref{fro}), the equality $Uf=0$ can
be written in the form:
\begin{equation}
\left(1-\frac{(1-x)^2}4\right)f\left(\frac{(1-x)^2}4\right)+
\left(1+\frac{(1-x)^2}4\right)
f\left(-\frac{(1-x)^2}4\right)=0,
\label{p11}
\end{equation}
$x\in[-1,1]$. For $x=-1$ the equality (\ref{p11}) implies
that $f(-1)=0$. Therefore $f(x)\equiv (1+x)g(x)$ for some
$g\in\A$. Substituting $(1+x)g(x)$ instead
of $f(x)$ into
(\ref{p11}), we obtain
\begin{equation}
\left(1-\frac{(1-x)^4}{16}\right)
\left(g\left(\frac{(1-x)^2}4\right)+
g\left(-\frac{(1-x)^2}4\right)\right)=0
\label{p12}
\end{equation}
for $x\in[-1,1]$. Denoting $y=(1-x)^2/4$, we arrive to
$g(y)=g(-y)$  for $y\in[0,1)$. Hence, $g$ is odd.

Suppose now that $f(x)\equiv(1+x)g(x)$ for odd $g\in\A$.
This implies the validity of (\ref{p12}) and
therefore (\ref{p11}), which is equivalent to $Uf=0$. Part
II of the theorem is proved.  Part I follows immediately
from parts~II and III.  \square

\section {Analytic continuation of eigenfunctions}

First, we introduce some notations, which we
shall use in Sections~3 and~4 without additional comments.
Let $\lambda$, $g$, and $f$ be as in the Proposition.
For $z\in\C$ and $r\in(0,+\infty)$ we denote
$$
D(z,r)=\{w\in\C:|w-z|<r\},\qquad
\D(z,r)=\{w\in\C:|w-z|\leq r\}.
$$
By $\C_-$ we denote the set $\C\setminus(-\infty,0)$.
We reserve the symbol $\sqrt z$ for the ``positive'' branch of
the square root on the set $\C_-$, i.e. $\sqrt z=\sqrt
re^{i\phi/2}$, where $z=re^{i\phi}$, $-\pi<\phi<\pi$.

For an infinite connected subset $A\subseteq\C$ we say that
a function $\phi:A\to\C$ is analytic if $\phi$ admits an
analytic extension to some open set, containing $A$. In
particular, $\A$ is the space of all functions analytic on
$[-1,1]$.  For a connected set $A\subseteq\C$ and a subset
$B\subseteq A$, having at least one limit point in $A$, we
say that a function $\phi:B\to\C$ is analytic on $A$ if $\phi$
admits a (unique) analytic extension on
 $A$.
We shall also denote by $\phi$ the extension.

We shall show that the function $f$ admits an analytic extension
to the disk $D(0,2+\sqrt 3)$ in several steps.

\lemma1 \\
Let $A$ and $B$ be subsets of $\C$, $\phi_1:A\to\C$
and $\phi_2: B\to\C$ be analytic functions, $M\subseteq
A\cap B$ be a set having at least one limit point in
$A\cap B$. Let also the set $A\cap B$ be connected
and $\phi_1\bigr|_M\equiv\phi_2\bigr|_M$. Then the function
\begin{equation*}
\phi:A\cup B\to\C,\qquad
\phi(z)=\left\{
\begin{array}{ll}\phi_1(z)&\ \ \text{if}\ \ z\in A;
\\ \phi_2(z)&\ \ \text{if}\ \ z\in B
\end{array}\right.
\end{equation*}
is well defined and analytic on $A\cup B$.
\epr

\proof Since $M$ has a limit point in the connected
set $A\cap B$, then according to the uniqueness theorem
\cite {titch} $\phi_1\bigr|_{A\cap
B}\equiv\phi_2\bigr|_{A\cap B}$.  Therefore $\phi$ is well
defined. Analyticity of $\phi$ follows from analyticity of
$\phi_1$ and $\phi_2$. \square

\lemma2 \\
The analyticity of $f$ on $\D(0,c)$
$(c\geq1)$ implies the analyticity of  $f$ on
$\D(0,c)\cup\D(1,2\sqrt c)$. \epr

\proof Let $f$ be analytic on $\D(0,c)$. Consider
$h:\D(1,2\sqrt c)\to\C$, $h(z)=(Uf(z)-g(z))/\lambda$.
Clearly $h$ is well-defined and analytic. Since
(\ref{fe1}) is valid on $[-1,1]$, we have that
$h(z)=f(z)$ for $z\in[-1,1]$. Lemma~1 implies that
the function
$$
q(y)=\left\{
\begin{array}{ll}f(y)&\ \ \text{if}\ \ y\in \D(0,c);
\\ h(y)&\ \ \text{if}\ \ z\in \D(1,2\sqrt c)
\end{array}\right.
$$
is well defined and analytic on $\D(0,c)\cup\D(1,2\sqrt
c)$. This is the desired analytic extension of $f$. \square

\lemma3 \\
The function $f$ is analytic on the closed disk
$\D(1,2)$ and the functional equation $(\ref{fe1})$ is
valid for all $x\in\D(1,2)$.  \epr

\proof It suffices to show that $f$ is analytic on
$\D(1,2)$ (the validity of the functional equation
$(\ref{fe1})$  for all $x\in\D(1,2)$ follows then from the
uniqueness theorem \cite {titch}). For this goal it
suffices to show that $f$ is analytic on
$\D(0,1)$. Analyticity on $\D(1,2)$  then follows
from Lemma~2. For $a\in(0,1]$ let
$$
K_a=\{z\in\D(0,1):|\im z|<a\}, \quad
\K_a=\{z\in\D(0,1):|\im z|\leq a\}.
$$

Suppose that $f$ is not analytic on $\D(0,1)$. Let us denote
$$
a=\sup\{b\in(0,1):f\text{\ is analytic on\ }K_b\}.
$$
Then $a\in(0,1]$, $f$ is analytic on $K_a$ and $f$ is not
analytic on $\K_a$.

Let $z\in\K_a$, $x=\re z$, $y=\im z$ and $w=(1-z)^2/4$.
Since $|z|\leq1$, we have that $|w|\leq(1+|z|)^2/4\leq1$.
Since $x^2+y^2\leq1$ and $|y|\leq a$ we have that $|\im
w|=|(x-1)y|/2\leq a|x-1|/2$. Therefore $|\im w|\leq a$.
Moreover $|\im w|=a$ if and only if $x=-1$ and $|y|=a$. But
then $|z|=\sqrt{1+a^2}>1$. Hence, $|\im w|<a$. Thus, we
have shown that
\begin{equation}
\pm(1-z)^2/4\in K_a\ \ \text{for any}\ \ z\in\K_a.
\label{ua}
\end{equation}

Formulas (\ref{ua}) and (\ref{fro}) imply that the function
$h(z)=(Uf(z)-g(z))/\lambda$ is well-defined and analytic on
$\K_a$. The equation (\ref{fe1}) implies that $f(x)=h(x)$
for $x\in[-1,1]$. Therefore $h$ is an analytic continuation
of $f$, i.e. $f$ is analytic on $\K_a$. This
contradiction completes the proof. \square

\lemma4 \\
{\bf I.} Let $c\in[2,+\infty)$,
$z\in\D(-1,c)\setminus D(1,c)$. Then
$1-2\sqrt{-z}\in\D(-1,c)$.

\noindent {\bf II.} Let $a\in[3,+\infty)$,
$z\in\D(0,a)$, $\re z\geq0$. Then $w,u\in D(0,a)$, where
$w=2\sqrt{z}-1$ and $u=z-w$.\epr

\proof I. According to the maximum principle~\cite {titch},
it suffices to show that
\begin{equation}
\sqrt{z}\in\D(1,c/2)\ \  \text{for any}\ \
z\in \partial M,
\label{I3}
\end{equation}
where $\partial M$ is the boundary of the set
$M=\{z\in\D(1,c):\re z\geq0\}$. Clearly (\ref{I3}) is
equivalent to
\begin{align}
&\sqrt{z}\in\D(1,c/2)\ \  \text{for any}\ \ z\in\Gamma_1
=\{it:t\in[-\sqrt{c^2-1},\sqrt{c^2-1}];
\label{I4}
\\
&\sqrt{z}\in\D(1,c/2)\ \  \text{for any}\ \ z\in\Gamma_2
=\{z\in\C:|z-1|=c,\ \re z\geq0\}.
\label{I5}
\end{align}

Parameterizing $z\in\Gamma_2$ by polar coordinates
$z=re^{i\phi}$, we obtain that (\ref{I4}) and
(\ref{I5}) are equivalent to
\begin{align}
&\max\{t+1-\sqrt{2t}:t\in[0,\sqrt{c^2-1}]\}\leq c^2/4;
\label{I6}
\\
&r+1\leq2\sqrt r\cos(\phi/2)+c^2/4
\text{\ \ if\ \ }
r^2-2r\cos\phi=c^2-1,\quad \phi\in[-\pi/2,\pi/2],
\label{I1}
\end{align}
respectively. Since $r^2-2r\cos\phi=c^2-1$, inequality from
(\ref{I1}) is equivalent to
$$
c^4/16+c^2(\sqrt r\cos(\phi/2)-1)\geq0.
$$

As $r$ and $\cos(\phi/2)$ are decreasing with respect to
$\phi\in[0,\pi/2]$ and the function $t+1-\sqrt{2t}$ on the
segment $[0,\sqrt{c^2-1}]$ takes the maximal value for
$t=\sqrt{c^2-1}$, we obtain that inequalities (\ref{I6})
and (\ref{I1}) are respectively equivalent to
\begin{align}
&4\alpha^4-8\alpha^2+8\alpha-3\geq0\text{\ \
for\ \ } \alpha\in [(3/4)^{1/4},+\infty);
\label{I7}
\\
&4\alpha^4+16\alpha-15\geq0\text{\ \
for\ \ } \alpha\in [(3/4)^{1/4},+\infty).
\label{I10}
\end{align}

Since $4\alpha^4-8\alpha^2+8\alpha-3=
(2\alpha^2+2\alpha-3)(2\alpha^2+2\alpha+1)$
and $4\alpha^4+16\alpha-15=
(2\alpha^2+2\alpha-3)(2\alpha^2-2\alpha+5)$,
the number $(\sqrt7-1)/2<(3/4)^{1/4}$ is the
maximal real zero of both polynomials
$4\alpha^4-8\alpha^2+8\alpha-3$ and
$4\alpha^4+16\alpha-15$. This proves (\ref{I7}) and
(\ref{I10}), which imply (\ref{I3}).

II. We have to prove that
$|w|<a$ and $|u|<a$. According to the maximum principle, 
it suffices to verify this for $z\in\Gamma$, where
$\Gamma=\Gamma_1\cup\Gamma_2$, $\Gamma_1=\{it:t\in[-a,a]\}$
and $\Gamma_2=\{ae^{i\phi}:\phi\in[-\pi/2,\pi/2]\}$.

For $z\in\Gamma_1$ we have
$$
|w|^2=4|t|-2\sqrt{2|t|}+1,\quad
|u|^2=t^2-2\sqrt2 |t|^{3/2}+4|t|-2\sqrt2 |t|^{1/2}+1.
$$
For $z\in\Gamma_2$ we have
\begin{align*}
|w|^2&=4a+1-4\sqrt a\cos(\phi/2),
\\
|u|^2&=a^2-4a^{3/2}\cos(\phi/2)+4a+2a\cos
\phi-4a^{1/2}\cos(\phi/2)+1.
\end{align*}

Differentiating these functions with respect to $t$ and
$\phi$, we find that both $|u|^2$ and $|w|^2$ for
$z\in\Gamma_1$ are maximal when $t=\pm a$ and that  both
$|u|^2$ and $|w|^2$ for $z\in\Gamma_2$ are maximal when
$\phi=\pm\pi/2$. Thus, in any case
\begin{equation}
|w|^2\leq 4a-2\sqrt{2a}+1\ \ \text{and}\ \
|u|^2\leq a^2-2\sqrt2 a^{3/2}+4a-2\sqrt2 a^{1/2}+1.
\label{xuy}
\end{equation}
Hence, it suffices to verify
that
\begin{equation}
a^2-4a+2\sqrt{2a}-1>0\text{\ \ and \ \ }
2\sqrt2 a^{3/2}-4a+2\sqrt2 a^{1/2}-1>0
\label{II3}
\end{equation}
for $a\geq3$. Both functions from (\ref{II3}) are
increasing for $a\geq3$. Hence, we should only prove
(\ref{II3}) for $a=3$, which is a simple arithmetic exercise.
\square

\lemma5 \\
Let $c\in[2,+\infty)$. Then the analyticity of $f$ on
$\D(1,c)$ implies the analyticity of $f$ on $\D(1,c)\cup
\D(-1,c)$. \epr

\proof Let $f$ be analytic on $\D(1,c)$. Pick
$\epsilon>0$ such that $f$ is analytic on
$\D(1,c+\epsilon)$ and let
\begin{equation}
\begin{array}{l}
S_0=\D(1,c+\epsilon)\cap(\D(1,c)\cup\D(-1,c)),
\\
S_{n+1}=S_n\cup\{z\in\D(-1,c):1-2\sqrt{-z}\in S_n\}
\\
{}\qquad=S_n\cup(\{-(1-w)^2/4:w\in S_n\}\cap\D(-1,c)).
\end{array}
\label{sn}
\end{equation}

Evidently $f$ is analytic on $S_0$.
It is easy to see that $S_n$, $n=0,1,2\ldots$, is an increasing
sequence of subsets of $\D(1,c)\cup\D(-1,c)$.

First, we shall show that analyticity of $f$ on $S_n$
implies analyticity of $f$ on $S_{n+1}$.  Let $f$ be
analytic on $S_n$. Consider the function $h:A_n\to\C$
\begin{equation}
h(z)=\frac{1+z}{z-1}f(-z)+
\frac{2\lambda}{1-z}f(1-2\sqrt{-z})+
\frac2{1-z}g(1-2\sqrt{-z}),
\label{hhh}
\end{equation}
where $A_n=\{z\in\D(-1,c):z\notin[0,+\infty),\
1-2\sqrt{-z}\in S_n\}$. Clearly $h$ is well defined and
analytic. Moreover, from (\ref{fe1}) and the
definition of $h$ it follows that
$h\bigr|_{(-1,0)}=f\bigr|_{(-1,0)}$.  Lemma~1 implies that
the function
$$
q:S_{n+1}=A_n\cup S_n\to\C,\quad q(z)=\left\{
\begin{array}{ll}f(z)&\ \ \text{if}\ \ z\in S_n;
\\ h(z)&\ \ \text{if}\ \ z\in A_n
\end{array}\right.
$$
is well defined and is an analytic extension of $f$ to $S_{n+1}$.
Therefore $f$ is analytic on $\ddd\bigcup_{n=0}^\infty
S_n$. It remains to show that
\begin{equation}
\D(1,c)\cup\D(-1,c)=\bigcup_{n=0}^\infty S_n.
\label{abc}
\end{equation}

From the definition of $S_n$, the point $z\in\D(-1,c)\setminus
D(1,c+\epsilon)$
belongs to $\ddd\bigcup_{n=0}^\infty S_n$, if and only if
there exists $m\in\N$ such that
$z_j\in\D(-1,c)\setminus D(1,c+\epsilon)$
for $0\leq j\leq m$ and
$z_{m+1}\in S_0$, where
\begin{equation}
z_0=z,\qquad z_{j+1}=1-2\sqrt{-z_j}.
\label{sz}
\end{equation}

Suppose that there exists $z\in\D(-1,c)\setminus
D(1,c+\epsilon)$ such that $z_n\in\D(-1,c)$  for all
$n\in\N$. Let $K$ be the closure of the set
$\{z_n:n=0,1,\dots\}$. Then $K$ is a closed subset of the
compact set $A=\D(-1,c)\setminus D(1,c+\epsilon)$, and
$\phi(K)\subseteq K$, where $\phi(z)=1-2\sqrt{-z}$. Since
$|\phi'(u)|<1$ for any $u\in A$, $\phi\bigr|_K:K\to K$
is a contraction. According
to the contraction map theorem, there exists a fixed point
of the map $\phi\bigr|_K$. But the unique solution (in
$\C\setminus[1,+\infty)$) of the equation $\phi(w)=w$ is
$w=-1\notin A$. This contradiction shows that for any
$z\in\D(-1,c)\setminus D(1,c+\epsilon)$ there exists the
first positive integer $m$ for which
$z_m\notin\D(-1,c)\setminus D(1,c+\epsilon)$. Then
$z_{m-1}\in\D(-1,c)\setminus D(1,c+\epsilon)$. According to
Lemma~4 $z_m=1-2\sqrt{-z_{m-1}}\in
\D(-1,c)\cap\D(1,c+\epsilon)\subset S_0$.
According to the description of the set
$\ddd\bigcup_{n=0}^\infty S_n$, we have that $\ddd
z\in\bigcup_{n=0}^\infty S_n$. This implies (\ref{abc}).
\square

\lemma6 \\
The function $f$ is analytic on the disk
$D(0,2+\sqrt3)$.  \epr

\proof Let $c_0=1$, $c_{n+1}=\sqrt{4c_n-1}$ for
$n=1,2,\dots$.  This sequence strictly increases and
converges to $2+\sqrt3$. From Lemma~3 it follows that $f$
is analytic on $\D(0,c_0)$. According to Lemmas~2 and~5
analyticity of $f$ on $\D(0,c_n)$ implies analyticity of
$f$ on $\D(1,2\sqrt{c_n})\cup \D(-1,2\sqrt{c_n})\supset
\D(0,c_{n+1})$. Therefore $f$ is analytic on $\D(0,c_n)$
for any $n=0,1,\dots$ Hence, $f$ is analytic on the set
$\ddd\bigcup_{n=0}^\infty\D(0,c_n)=D(0,2+\sqrt3)$. \square

\section {Proof of the Proposition}

I. Without loss of generality we can assume that
$g(-1)=0$ and $f(-1)=0$. If this is not the case, we can achieve
these conditions just by adding suitable constants to $f$ and $g$.
Therefore $f(x)=\phi(x)(1+x)$, $g(x)=\psi(x)(1+x)$ and
$\psi\in\E$, $\phi\in\A$. Moreover analyticity of $f$ on
a connected set $A\supset[-1,1]$ implies  analyticity of
$\phi$ on the same set $A$. Let
\begin
{equation}
\begin{array}{ll}
f_0(x)=(\phi(x)+\phi(-x))/2,&\quad f_1(x)=(\phi(x)-\phi(-x))/2;
\\
g_0(x)=(\psi(x)+\psi(-x))/2,&\quad g_1(x)=(\psi(x)-\psi(-x))/2.
\end{array}
\label{evod1}
\end{equation}

Then $g_0,g_1\in\E$, $f_0,f_1\in\A$. The analyticity of $f$
on a connected symmetric (with respect to 0) set
$A\supset[-1,1]$ implies the analyticity of $f_0,f_1$ on the
same set. Thus, according to Lemma~6 $f_0$ and $f_1$ are
analytic on $D(0,2+\sqrt3)$.

Evidently $f_0,g_0$ are even, $f_1,g_1$ are odd and
\begin{equation}
f(x)=(1+x)(f_0(x)+f_1(x)),\quad
g(x)=(1+x)(g_0(x)+g_1(x)).
\label {evod2}
\end{equation}

From (\ref{fe1}) and (\ref{evod2}) it follows that
$$
\left(1-\frac{(1-x)^4}{16}\right)
f_0\left(\frac{(1-x)^2}{4}\right)=
\lambda(1+x)(f_0(x)+f_1(x))+(1+x)(g_0(x)+g_1(x))
$$
for any $x\in[-1,1]$. Dividing by $(1+x)$ we obtain
\begin{equation}
\frac1{16}(15-11x+5x^2-x^3)
f_0\left(\frac{(1-x)^2}{4}\right)=
\lambda(f_0(x)+f_1(x))+g_0(x)+g_1(x).
\label{fe3}
\end{equation}
Adding (\ref{fe3}) for $x$ with (\ref{fe3}) for $-x$ we
obtain that for any $x\in[-1,1]$,
\begin{equation}
\begin{array}{l}\ddd
\frac1{32}(15-11x+5x^2-x^3)
f_0\left(\frac{(1-x)^2}{4}\right)+
\\
{\qquad}\ddd+\frac1{32}(15+11x+5x^2+x^3)
f_0\left(\frac{(1+x)^2}{4}\right)=
\lambda f_0(x)+g_0(x).
\end{array}
\label {fe4}
\end{equation}

Let us prove that $f_0\in\E$. Suppose that $f_0\notin\E$.
Since $f_0$ is analytic on $D(0,2+\sqrt3)$, there exists
$a\in[2+\sqrt3,+\infty)$ such that $f_0$ is analytic on
$D(0,a)$ and is not analytic on $\D(0,a)$. Since $f_0$ is
even, we have that $f_0$ is not analytic on the set
$B=\{z\in\D(0,a):z\neq0,\re z\geq0\}$.
According to
Lemma~4 $x=2\sqrt y-1\in D(0,a)$ and $y-x\in D(0,a)$
for any $y\in B$.
Since $15+11x+5x^2+x^3\neq0$ for $y\in B$, the function
$$
h(y)=\frac{(x^3-5x^2+11x-15)f_0(y-x)+32\lambda f_0(x)+32g_0(x)}
{15+11x+5x^2+x^3},
$$
where $x=x(y)=2\sqrt y-1$, is well defined and analytic on
$B$. On the other hand, (\ref{fe4})
and the definition of $h$ imply that $h(x)=f(x
)$ for all
$x\in(0,1)$. The uniqueness theorem implies
that $h$ is an analytic extension of $f_0$ from
$(0,1]$ to $B$, which does not exist. This contradiction
proves that $f_0\in\E$.

From (\ref{fe3}) it follows that
\begin{equation}
f_1(x)=
\frac1{16\lambda}(15-11x+5x^2-x^3)
f_0\left(\frac{(1-x)^2}{4}\right)
-f_0(x)-\frac1\lambda(g_0(x)+g_1(x)).
\label {F1}
\end{equation}
Therefore $f_1\in\E$. Formula
(\ref{evod2}) implies that $f$ is entire and Part~I is proved.

\smallskip

II. Let $g\in\P$. We have to prove that $f\in\P$. According
to (\ref{evod1}) $g_0,g_1\in\P$. Let $k=\deg g_0$ if
$g_0\not\equiv0$ and $k=0$ if $g_0\equiv0$. Let us show
that  $f_0\in\P$ and $\deg f_0\leq k/2-1$.

The functional equation (\ref{fe4}) can be
rewritten in the form
\begin{equation}
\begin{array}{l}
\ddd f_0\biggl(\frac{(1+x)^2}{4}\biggr)
=\frac{32\lambda f_0(x)}{x^3+5x^2+11x+15}+
\frac{32 g_0(x)}{x^3+5x^2+11x+15}+
\\
{}\quad\ddd+
\frac{x^3-5x^2+11x-15}{x^3+5x^2+11x+15}
f_0\biggl(\frac{(1-x)^2}{4}\biggr).
\end{array}
\label{M}
\end{equation}

Since $f_0$ is even
\begin{equation}
M(R)=\max_{|x|=R}|f_0(x)|=\max_{|x|=R,\ \re x\geq0}|f_0(x)|.
\label{mmm}
\end{equation}
Let $y\in\C$, $|y|=R$, $\re y\geq0$, $x=2\sqrt{y}-1$,
$w=y-x$. Then $|x|=2\sqrt R+O(1)$ and
$$
|w|=|y|\biggl|1-\frac
yx\biggr|=R\biggl|1-\frac2{\sqrt y}+\frac1y\biggr|
=R\biggl|1-\frac2{\sqrt y}\biggr|+O(1).
$$
All $O$-symbols are considered here for $R\to\infty$.
The number $|1-(2/\sqrt y)|$ for $|y|=R$, $\re y\geq0$
is maximal for $y=\pm Ri$. Therefore
$$
|w|\leq R\biggl|1-\frac{\sqrt2(1+i)}{\sqrt R}\biggr|+O(1)=
R-\sqrt {2 R}+O(1)
$$
and
\begin{equation}
|w|<R-\sqrt R,\qquad |x|=2\sqrt R+O(1)<R-\sqrt R
\label{est1}
\end{equation}
for sufficiently large $R$. Formula
(\ref{M}) implies that
\begin{equation}
f_0(y)=\frac{32\lambda f_0(x)}{x^3+5x^2+11x+15}+
\frac{32 g_0(x)}{x^3+5x^2+11x+15}+
\frac{x^3-5x^2+11x-15}{x^3+5x^2+11x+15}f_0(w).
\label{MM}
\end{equation}
Note that
$$
\left|\frac{x^3-5x^2+11x-15}{x^3+5x^2+11x+15}
\right|=\left|1-\frac{10}{x}+O\left(\frac1{x^2}\right)\right|
=\left|1-\frac{5}{\sqrt y}\right|+O(R^{-1}).
$$
The number $|1-(5/\sqrt y)|$ for $|y|=R$, $\re y\geq0$
is maximal for $y=\pm Ri$. Therefore
\begin{equation}
\left|\frac{x^3-5x^2+11x-15}{x^3+5x^2+11x+15}
\right|\leq 1-\frac5{\sqrt{2R}}+O(R^{-1}).
\label{est2}
\end{equation}
Formulas (\ref{MM}), (\ref{mmm}), (\ref{est1}) and
(\ref{est2}) imply that
$$
M(R)\leq M(R-\sqrt R)\left(1-
\frac5{\sqrt{2R}}+O(R^{-1})\right)+O(R^{(k-3)/2}).
$$

If $f_0$ is a polynomial of degree at most $k/2-1$, we have proved
the statement. Otherwise $R^{k/2}=O(M(R))$. Hence
$R^{(k-3)/2}=O(M(R-\sqrt R)/R)$, and
\begin{equation}
M(R)\leq M(R-\sqrt R)\left(1-
\frac5{\sqrt{2R}}+O(R^{-1})\right).
\label{otherw}
\end{equation}
Therefore $M(R)\leq M(R-c\sqrt R)$ for sufficiently large $R$.
Hence $M(R)=O(1)$ and $f_0$ is a constant according to
the Liouville theorem \cite{titch}. Hence $M(R)$ is
constant. Formula (\ref{otherw}) implies then that
$M(R)\equiv0$. Therefore $f_0\equiv0$ and $k=0$ according
to (\ref{M}). Thus, anyway $f_0\in\P$ and $\deg f_0\leq
k/2-1$.

From~(\ref{F1}) we find that $f_1\in\P$ and
$$
\deg f_1 \leq \max\{k+1,\deg g_1\}.
$$
Then using~(\ref{evod2}) we find that $f\in\P$ and
$\deg f \leq \deg g + 1$.

III. Suppose that $g\in\P$
and $f\in\A$ satisfies (\ref{fe1}). According
to Part~II of the Proposition $\ddd
f(x)=\sum_{k=0}^na_kx^k$, $a_n\neq0$. If $n$ is odd then
according to (\ref{fro}) $\deg Uf=2n+2$ and $\deg g=\deg
(Uf-\lambda f)=2n+2$ is even. If $n$ is even and $n\neq\deg Uf$,
then $\deg g=\max\{\deg f,\deg Uf\}$ is
even. It remains to consider the case $\deg Uf=n$. Since
$\deg Uf$ is alwais a multiple of $4$, we have that $n=4m$,
$m=0,1,\dots$.  According to (\ref{fro})
\begin{equation}
Uf(x)=\sum_{l=1}^{2m}(a_{2l}-a_{2l-1})2^{-4l}(1-x)^{4l}+a_0.
\label{xzxz}
\end{equation}
Since $\deg Uf=4m$ we have that $a_{2l}=a_{2l-1}$ if
$m<l\leq2m$ and $a_{2m}\neq a_{2m-1}$. Substituting
(\ref{xzxz}) into (\ref{fe1}) and taking these
relations into account, we obtain that $\deg g=4m$ or $\deg
g=4m-1$.  In any case $\deg g$ does not have form $4j+1$.

\section {Proof of Theorem 3}

We start with three lemmas.

\lemma7 Let $q\in(0,1)$ and $z\in\D(q,1+q)$. Then
$(1-z)^2/4\in D(q,1+q)$. \epr

\proof We have to prove that
\begin{equation}
|(z-1)^2-4q|<4+4q.
\label{hotimm}
\end{equation}
Since $z\in\D(q,1+q)$, we find that $z=q+(1+q)u$, where
$|u|\leq1$. Obviously (\ref{hotimm}) is
equivalent to
\begin{equation}
|1-6q+q^2-2(1-q^2)u+(1+2q+q^2)u^2|<4+4q.
\label{hotim1}
\end{equation}

If $1-6q+q^2\geq0$, we have
$$
|1-6q+q^2-2(1-q^2)u+(1+2q+q^2)u^2|\leq
1-6q+q^2+2(1-q^2)+(1+2q+q^2)=4-4q<4+4q.
$$

If $1-6q+q^2<0$, we have
\begin{align*}
&|1-6q+q^2-2(1-q^2)u+(1+2q+q^2)u^2|\leq
-1+6q-q^2+2(1-q^2)+(1+2q+q^2)
\\
&\qquad=4+4q-2(1-q)^2<4+4q .
\end{align*}
\square

\lemma8 \\
Let $\H$ be the Hardy space $H^2$ in the upper half-plane
$\Pi=\{z\in\C:\re z>\alpha\}$, $\alpha\in\R$,
$\nu,\phi:\overline\Pi\to\C$ be bounded analytic functions
and $c,\epsilon\in(0,+\infty)$ be such that

\begin{align}
&\re\phi(z)\geq\epsilon\text{\ for all }z\in\Pi;
\label{epep}
\\
&\int\limits_{-\infty}^{+\infty}|\nu(\alpha+is)|^2\,ds\leq A
<\infty;\quad
\int\limits_{-\infty}^{+\infty}|\phi(\alpha+is)-c|^2\,ds\leq A
<\infty.
\label{intint}
\end{align}

Then the operator $S:\H\to\H$,
$Sf(z)=(1+\nu(z))f(z+\phi(z))$ is the sum of two operators
$A$ and $B$, where $A$ is bounded, self-adjoint and has
purely absolutely continuous spectrum $[0,1]$ and $B$ is a
Hilbert--Schmidt operator. \epr

\proof Without loss of generality we assume $\alpha=0$.
According to the Paley--Wiener theorem
(\cite{hoffman}, Chapter~8), the
Laplace transform
$$
Lg(z)=\frac1{\sqrt{2\pi}}\int_0^\infty e^{-zt}g(t)\,dt
$$
is a unitary operator from $L_2[0,+\infty)$ onto $\H$.

Let ${\cal D}_+$ be the subspace of $L_2[0,+\infty)$,
consisting of infinitely differentiable  functions
with compact support lying in $(0,+\infty)$.
Consider the operator $\widetilde
S=L^{-1}SL:L_2[0,+\infty)\to L_2[0,+\infty)$.
Using the standard formula for the inverse Laplace
transform we obtain that
$$
\widetilde S g(t)=\frac1{2\pi}\int\limits_{-\infty}^{+\infty}
\int\limits_0^{+\infty} g(\tau)e^{is(t-\tau)}e^{-\phi(is)\tau}
(1+\nu(is))\,d\tau\,ds
$$
for any $g\in{\cal D}_+$. Therefore
\begin{equation}
\widetilde S g(t)=g(t)e^{-ct}+\int\limits_0^{+\infty} g(\tau)
K(\tau,t)\,d\tau,
\label{tilt}
\end{equation}
where
$$
K(\tau,t)=\frac1{2\pi}\int\limits_{-\infty}^{+\infty}
e^{is(t-\tau)}\left[e^{-\phi(is)\tau}
(1+\nu(is))-e^{c\tau}\right]\,ds.
$$
The existence of the last integral (in the principle value
sense) for any $\tau$ and almost all $t$, follows from the
Plancherel theorem, since formula (\ref{intint}) implies
square integrability, with respect to $s$, of the function
$e^{-is\tau}\left[e^{-\phi(is)\tau}
(1+\nu(is))-e^{c\tau}\right]$.

According to the Parseval identity and using formulas
(\ref{epep}) and (\ref{intint}), we obtain
\begin{align*}
&\int\limits_{-\infty}^{+\infty}|K(\tau,t)|^2dt=\frac1{2\pi}
\int\limits_{-\infty}^{+\infty}\left|e^{-\phi(is)\tau}
(1+\nu(is))-e^{c\tau}\right|^2ds\leq
\\
&\leq\frac{e^{-2\tau\epsilon}}{\pi}\int\limits_{-\infty}^{+\infty}
\left(
\left|e^{-(\phi(is)-\epsilon)\tau}-e^{(c+\epsilon)\tau}
\right|^2+\left|e^{-(\phi(is)-\epsilon)\tau}\nu(is)
\right|^2\right)\,ds\leq
\\
&\leq\frac{e^{-2\tau\epsilon}}{\pi}\int\limits_{-\infty}^{+\infty}
\left(\tau^2|\phi(is)-c|^2+|\nu(is)|^2\right)\,ds\leq
\frac{Ae^{-2\tau\epsilon}(\tau^2+1)}{\pi}.
\end{align*}

According to Fubini's theorem
$$
\int\limits_{0}^{+\infty}\int\limits_{0}^{+\infty}|K(\tau,t)|^2dtd\tau
\leq\int\limits_{0}^{+\infty}\int\limits_{-\infty}^{+\infty}
|K(\tau,t)|^2dtd\tau\leq\frac A\pi\int\limits_{0}^{+\infty}
e^{-2\tau\epsilon}(\tau^2+1)\,d\tau<+\infty.
$$

Therefore, $K\in L_2([0,+\infty)^2)$. Since ${\cal D}_+$ is
dense in $L_2[0,+\infty)$, formula
(\ref{tilt}) is valid for any $g\in L_2[0,+\infty)$.
Therefore $\widetilde S=\widetilde A+\widetilde B$, where
$$
\widetilde Ag(t)=g(t)e^{-ct}\text{ \ and \ }\widetilde
Bg(t)=\int\limits_0^{+\infty} g(\tau)K(\tau,t)\,d\tau,
$$
the operator $\widetilde A$ is bounded, selfadjoint and has
purely absolutely continuous spectrum $[0,1]$, and the
operator $\widetilde B$ is a Hilbert--Schmidt operator.
It remains to note that operators $S$ and $\widetilde S$ are
unitarily equivalent. \square

The following lemma is a consequence of Lemma~5.2 of
Chapter~1 in the book \cite {g-k}. \rm

\lemma9 \\ Let $X$ be a Banach space, $A$ be a
bounded linear operator on $X$ with simply connected
spectrum $\sigma(A)=K$ without isolated points and $B$ be a
compact linear operator on $X$. Then the spectrum of $A+B$
is the union of $K$ and a finite or countably infinite set of 
isolated eigenvalues of finite multiplicity. \epr

Now we can prove Theorem~3. Evidently $U=U^0+U^1$, where
\begin{align}
U^0f(x)&=\frac12\left(1+\frac{(1-x)^2}4\right)
f\left(-\frac{(1-x)^2}4\right),
\label {u0}
\\
U^1f(x)&=
\frac12\left(1-\frac{(1-x)^2}4\right)
f\left(\frac{(1-x)^2}4\right).
\label {u1}
\end{align}

According to Lemma~7, there exists a $r>1+q$
such that for any $f\in X$ the function $U^1f$ admits an
analytic extension to $\D(q,r)$. Let $X_r$ be the Hardy
space in the disk $D(q,r)$. Then formula (\ref{u1}) defines
a bounded linear operator $U^1_r$ from $X$ to $X_r$. The
operator $U^1_X:X\to X$ is the superposition of
$U_r^1$ and the identity embedding $J$ of $X_r$ into $X$.
Since the operator $J$ is nuclear ($J$ has the $s$-numbers
$(1+q)^nr^{-n}$)), we find that $U^1_X=U^1\bigr|_X:X\to X$
is also nuclear and therefore compact.

Let $\H$ be the Hardy space in the upper half-plane 
$\{z\in\C:\re z>(1+q)^{-1}\}$. One can easily verify that the 
operator $M:X\to\H$,
\begin{equation}
\label {M-def}
Mf(z)=\frac1z f\biggl(\frac2z-1\biggr),
\end{equation}
is unitary up to a multiplication on a positive constant.
Then the operators $U^0_X$ and $W^0=MU^0M^{-1}:\H\to\H$ 
are unitarily equivalent.

From the definitions of (\ref{u0}), (\ref{M-def}) we obtain that
$$
M^{-1}f(z)=\frac2{z+1}g\biggl(\frac2{z+1}\biggr)\ \
\text{and}\ \  W^0f(z)=(1+\nu(z))f(z+\phi(z)),
$$
where
$$
\nu(z)=\frac{1-z}{2z^2-z}\text{\ \ and\ \ }
\phi(z)=\frac12+\frac1{4z-2}.
$$

One can easily verify that $\phi$ and $\nu$ satisfy all
conditions of Lemma~8 with $\epsilon=c=1/2$. According to
Lemma~8, $W^0$ is a sum of a self-adjoint operator with the
purely absolutely continuous spectrum $[0,1]$ and a 
Hilbert--Schmidt operator. Since $U^0_X$ and $W^0$ are unitarily 
equivalent and $U^1_X$ is nuclear, we have that $U_X$ is a sum 
of a self-adjoint operator with purely absolutely continuous
spectrum $[0,1]$ and a Hilbert--Schmidt
operator. It remains to apply Lemma~9. \square

\section {Concluding remarks}

1. So far there exist very few results on the spectral properties of
the F.P.O. of the maps with parabolic neutral fixed points. We
would like to point out the result of H.~Rugh \cite {rugh}, who
considered the F.P.O. of piecewise analytic maps, which are expanding
everywhere except one parabolic fixed point. Namely, he constructed
a specific map-dependent Banach space of analytic functions, where
the spectrum of the F.P.O. consists of the segment [0,1] and some
isolated normal eigenvalues. This space is in fact the image of
$L_1[0,+\infty)$ with respect to some map-dependent integral
transformation similar to the Laplace transform.

The cusp map does not satisfy the conditions
of Rugh's theorem because of the cusp-shaped
singularity. Nevertheless, we proved that the F.P.O. of the
cusp map has similar spectral properties in the Hardy spaces
$H^2$ in the disks $D(q,1+q)$, $0<q<1$.
We also conjecture that the spectrum of the F.P.O. $U$
of the cusp map in the Hardy spaces $H^2$ in the disks $D(q,1+q)$,
$0<q<1$ is precisely the segment $[0,1]$, i.e.,
the set of isolated eigenvalues of $U$ is empty.
Note that the functions of these Hardy spaces as well as the
functions of Rugh's spaces are analytic in all points of
the segment except at the parabolic fixed point ($x=-1$ in
the case of the cusp map). However, we should notice that
the spectrum of the F.P.O. of a map $S$ in spaces of
analytic functions with singularity at a fixed point of $S$
may differ considerably from the spectrum in spaces of
everywhere analytic functions. We have proved that this is
precisely the case for the cusp map.

2. The theory of the point spectrum for the maps has been
recently developed in terms of locally convex topological vector
spaces~\cite {ant76}. For different classes of observables
the same evolution law may have different resonances i.e.
different rates of approach to equilibrium. However, once the class of
observables is chosen, the resonance structure is
unique~\cite {ant76,ant94}. In terms of the assumptions of~\cite {ant76},
the admissible point spectra for a given map are described. 
Here we see that for the cusp map, we have continuous spectra in Hardy
spaces. This type of spectra were not addressed in~\cite {ant76}.

3. We would like also to notice that Theorems~1 and 2 remain valid if
one replaces the F.P.O.~(\ref{fro}) of the cusp map by some positive
transfer operator~\cite {baladi} of the cusp map, for example:
$$
\tilde Uf(x)=\frac12 f\left( \frac{(1-x)^2}{4}\right)+
\frac12 f\left(-\frac{(1-x)^2}{4}\right).
$$
Of course, Theorems~1 and 2 do not remain valid for all positive
transfer operators in the class considered in~\cite {baladi}.
For example, let us consider the operator
$$
Wf(x)=\left(\frac12-\beta+\frac{a(x)}{4}\right)f(a(x))+
\left(\frac12+\beta-\frac{a(x)}{4}\right) f(-a(x)),
$$
where $a(x)=(1-x)^2/4$. For the real parameter $\beta\in[-1/4,1/2]$,
this is a positive transfer operator for the cusp map, and $W1=1$.
On the other hand, the function
$$
f(x)=x-\frac{x^2}2+\frac{\beta}{2(1-\beta)}
$$
is the eigenfunction of $W$ corresponding to the eigenvalue $\beta$:
$Wf=\beta f$. Hence Theorems~1 and 2 are not valid for the operator $W$.

4. There are few questions which remain open for the cusp map.
First, the question about the asymptotics of the autocorrelation
function. As the eigenvalues of the F.P.O. of the
family~(\ref{familydefinition}) tend to unity when
$\epsilon\rightarrow 1/2$, one
can expect non-exponential decrease of the autocorrelation function.
The estimations in~\cite {lust} show that the autocorrelation
function $C(n)$ decreases as $1/n$, when $n\to\infty$. However, this
conjecture has not yet been analytically proven. Another question 
addresses the choice of the space of analytic functions where the 
spectrum of the F.P.O. is naturally defined by the dynamics of the 
map.

\vskip 5mm

{\bf Acknowledgments.} We would like to thank Profs. Ilya Prigogine
and Victor Sadovnichy for helpful discussions. We would like also 
to thank the referee who motivated us to formulate and prove Theorem~3
which added value to the paper.
This work enjoyed the financial support of the European Commission,
project IST-2000-26016 IMCOMP, and the Belgian Government through
the Interuniversity Attraction Poles. S.~A.~Shkarin is
supported by the Alexander von Humboldt foundation.

\end {document}